\documentclass[%
 reprint,
groupedaddress,
showpacs,preprintnumbers,
 amsmath,amssymb,
 aps,
]{revtex4-1}

\usepackage{amsmath}
\usepackage{graphicx}
\usepackage{dcolumn}
\usepackage{bm}

\usepackage[utf8]{inputenc}
\usepackage[caption=false]{subfig}
\usepackage{floatrow}
\usepackage{float}
\usepackage[rgb]{xcolor}
\usepackage{epsfig}
\usepackage{tabularx}

\usepackage{capt-of}
\usepackage{floatrow}

\begin{document}


\title{Deep Learning Super-Diffusion in Multiplex Networks}

\thanks{All supporting data and source code are available online.}

\author{Vito M.~Leli}
 \altaffiliation{Skolkovo Institute of Science and Technology\\ 3 Nobel Street, 
Moscow, Russia 121205}
\author{Saeed Osat}
\author{Timur Tlyachev}
\author{Dmitry V.~Dylov}\email{d.dylov@skoltech.ru}
\author{Jacob D.~Biamonte}

\date{\today}

\begin{abstract}
Complex network theory has shown success in understanding the emergent and collective behavior of complex systems~\cite{Newmanbook}.   Many real-world complex systems were recently discovered to be more accurately modeled as multiplex networks~\cite{GinestraBook, BOCCALETTI20141,Lee2015, Kivela2014,MultNetMathPRX}---in which each interaction type is mapped to its own network layer; e.g.~multi-layer transportation networks, coupled social networks,  metabolic and regulatory networks, etc. A salient physical phenomena emerging from multiplexity is super-diffusion: exhibited by an accelerated diffusion admitted by the multi-layer structure as compared to any single layer. Theoretically super-diffusion was only known to be predicted using the spectral gap of the full Laplacian of a multiplex network and its interacting layers. Here we turn to machine learning which has developed techniques to recognize, classify, and characterize complex sets of data. We show that modern machine learning architectures, such as fully connected and convolutional neural networks, can classify and predict the presence of super-diffusion in multiplex networks with 94.12\% accuracy. Such predictions can be done {\it in situ}, without the need to determine spectral properties of a network. 
\end{abstract}

                              
\maketitle

\nocite{*} 

Complex systems are often well-approximated by complex networks or graphs, i.e.~elements of the system can be represented as nodes where connections between nodes correspond to non-trivial interaction patterns~\cite{Newmanbook, EstradaBook}. A multiplex (multi-layer) network is a generalized framework to study complex networks with more than one interacting layer: each layer represents a different type of interaction between constituent nodes~\cite{GinestraBook, BOCCALETTI20141,Lee2015, Kivela2014,MultNetMathPRX}. Vast efforts have only more recently been devoted to study the differences and new phenomenon in networks when multiplexity is taken into account \cite{Szell2010,Mucha2010}. By considering multiplexity, a network becomes a so called multi-layer structure---thereby aggregating all layers into one single-layer network, or considering each layer separately typically entails a lossy compression~\cite{MultNetMathPRX}.

Considering multiplexity has lead to the discovery of new behavior in networked systems. The seminal paper by Buldyrev~\cite{Buldyrev2010} proves multiplex networks (interdependent networks in their case) exhibit a first order abrupt phase transition in their percolation phase diagram. This finding is in contrast to continuous phase transitions exhibited in the percolation phase diagram of monoplex networks (single-layer networks)~\cite{Cohen2000PRL,Newman2003Siam}. Then numerous papers tried to understand different behaviors that arise  in studying multilayer graphs. Examples include, structural processes such as site, bond percolation~\cite{Hackett2016PRX,Bianconi2016PRE, Cellai2013PRE,Radicchi2015NatPhys,RadicchiBianconi2017PRX,ghavasieh2020unraveling}, first-depth percolation or observability, kcore~\cite{AzimiKcore2014PRE}, optimal percolation~\cite{Osat2017NatComm}, as well as dynamical processes such as epidemic spreading~\cite{DickisonEpidemic2012PRE}, synchronization~\cite{Genio2016SciAdv}, diffusion~\cite{DeDomenico2014PNAS,GomezSupDiff2013PRL,Tejedor2018PRX}, controllability~\cite{Posfai2016PRE} and recently multi-dynamics running on top of multilayer structures~\cite{WANG2015LIFE,Granell2013PRL,Granell2014PRE, Lima2015, AzimiEpidemic2016PRE}.   Here we focus on super-diffusion---a physically observed phenomena in which a coupling between two distinct types of networks, creates `shortcuts' e.g.~using a bus and then a  metro can result in crossing town much faster than relying on either buses or the metros alone.

Conventionally, multiplex networks and particularly super-diffusion has been studied using a variety of tools from statistical physics, spectral graph theory and other tailored methods which are paired with powerful computer simulations and/or numerical methods~\cite{GinestraBook}. The computer simulations are case dependent; e.g.~Monte Carlo sampling is used to study percolation, or simulation of epidemic spreading, whereas spectral methods are related to some specific problems including community detection, percolation threshold, combinatorial problems, etc. A  specific (yes/no) question answered using spectral graph theory is the presence of super-diffusion in a given multiplex network. In current approaches, the spectral gap of the graph Laplacian---the first non-trivial eigenvalue of the Laplacian---is calculated to determine whether a multiplex system exhibits super-diffusion or not~\cite{GomezSupDiff2013PRL}. 

Machine learning has been successfully applied in a number of complex networks studies \cite{MLCompNet}. This includes hyperbolic embedding of complex networks~\cite{Muscoloni2017NatComm}, link prediction~\cite{Hasan06linkprediction} and representation learning~\cite{Hamilton2017RepresentationLO}.  However, augmenting the tool-set of complex networks with modern deep learning approaches is still in its nascent stages, with work reporting link prediction \cite{LinkML2018}. The ability of modern machine learning techniques to classify, identify, and/or interpret massive data sets such as images foreshadows their suitability to provide network scientists with similar success cases when studying the extremely large data sets embodied in the state-space of complex networks.  

In this study, we employ a deep learning approach which trains a neural network to (successfully) classify the presence of super-diffusion phenomena in multiplex networks. We represent the data-set in a gray-scale image and employ two different neural network architectures from machine learning (multilayer perceptron) and deep learning (convolutional neural network~\cite{goodfellow2016deep}, \cite{lecun1998gradient}). Definitively, our trained neural networks predict super-diffusion in multiplex networks with 94.12\% accuracy. Such predictions can be done in situ, without the need of a networks spectral properties. We believe that the success of this finding will foster a wide application of deep learning to study complex networks. 

\paragraph{Super-diffusion.}
Diffusion processes are among the simplest dynamics studied on multiplex structures~\cite{GomezSupDiff2013PRL}. 
Gomez et al.~\cite{GomezSupDiff2013PRL} studied diffusion in a duplex (two layer multiplex network) and found out that under certain circumstances, diffusion can be faster in a multiplex network compared to diffusion in each of the layers  separately~\cite{GomezSupDiff2013PRL}. 
This phenomena---which is called super-diffusion in multiplex networks---can be predicted by leveraging the relation between the spectrum of a Laplacian matrix and diffusion time-scales. The combinatorial Laplacian (as called in graph theory) governs dynamics of diffusion running on top of the graph i.e. the Laplacian is the generator of the stochastic process describing the random walk on the graph. While there are fundamental connections between the spectrum of the Laplacian of a graph with structural properties of a graph itself, there are also relationships between the spectrum and dynamics running on top of the graph. In the case of diffusion processes, the second smallest eigenvalue (the first nontrivial one) of the Laplacian controls the timescale of the diffusion process. In the case of multiplex networks, the Laplacian matrix has the counterpart which is called supra-Laplacian~\cite{MultNetMathPRX} which is the same as the Laplacian matrix yet offers an easier interpretation (with proper labeling of nodes) as a block diagonal matrix in which diagonal blocks are the Laplacian of each of the layers (capturing intra-layer connections) and off-diagonal blocks which capture interlayer connections. 
\begin{equation}
\mathcal{L}(\mathcal{L}_1, \mathcal{L}_2, w)=\left(\begin{array}{cc} \mathcal{L}_1 + w\openone & -w\openone\\ 
-w\openone & \mathcal{L}_2 +  w\openone \end{array}\right)
\end{equation}
Gomez et al.~\cite{GomezSupDiff2013PRL} showed that strength of interlayer connections ($w$) is a tunable parameter between slow and fast diffusion regimes. For weak inter-layer connections, multiplexity slows down the diffusion while strong inter-layer connections define a new behavioral regime which asymptotically approaches the behavior of the aggregated layer (the superposition of the two layers). In the strong coupling regime, if 
\begin{equation}
 \lambda_2(\mathcal{L}) \geq \text{Max}\{ \lambda_2(\mathcal{L}_1), \lambda_2(\mathcal{L}_2)\}
\end{equation}
super-diffusion is exhibited or in other words, multiplexity helps diffusion happen faster. These two distinct regimes are direct consequence of structural transitions in the formation of multiplex networks starting from separate layers and then placing inter-layer links between them and increasing the weight of these links~\cite{Radicchi2013NatPhys,Radicchi2018PRX, Rapisardi2018PRE, Darabi2015PRE, MARTINHERNANDEZ201492}. The phenomena was also reported recently for directed multiplex networks~\cite{Tejedor2018PRX}.

\begin{figure}[H]
  \includegraphics[width=1\textwidth]{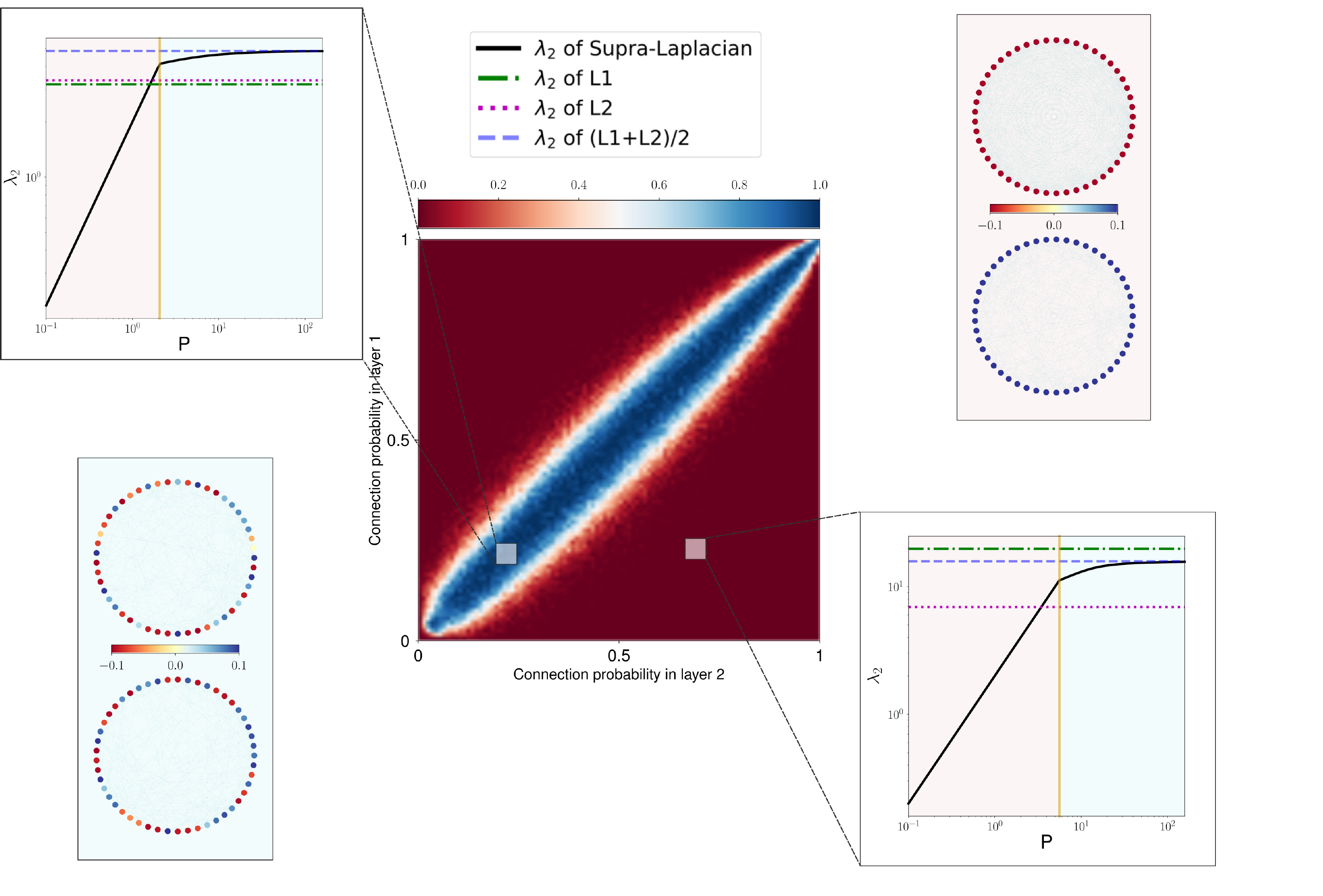}
  \caption{({\bf Super-Diffusion in multiplex networks.})
  A heat map of super-diffusion in the whole phase diagram of 2-layer multiplex networks was created, in which each layer has been formed using the Erd\H{o}s-R\'enyi random graph model $\mathcal{G}(N,p)$, where $N$ is number of  nodes ($N=50$) and $p$ is the connection probability. For each ($p_1,p_2$) we create 10 samples of the 2-layer random multiplex networks, then we check the condition of super-diffusion;
$\lambda_2(\mathcal{L}) \geq \text{Max}\{ \lambda_2(\mathcal{L}_1), \lambda_2(\mathcal{L}_2)\}$
and then color code the probability of the super-diffusion for the bin ($p1,p2$). The plot indicates that super-diffusion happens mainly in the region that connections probabilities are close to each other. The left (right) zoomed window shows the whole phase diagram of $\lambda_2(\mathcal{L})$ for the case that super-diffusion happens (does not happen). Vertical yellow line in the zoomed windows represents the phase transition point: ($p^*=\frac{1}{2}\lambda_2(Q)$ where $Q=\mathcal{L}_1\bar{\mathcal{L}}^{\dagger}\mathcal{L}_2$ and $\bar{\mathcal{L}}^{\dagger}$ is the Moore-Penrose pseudoinverse of $\bar{\mathcal{L}}=\frac{\mathcal{L}_1+\mathcal{L}_2}{2}$) \cite{Darabi2015PRE}. The circular layout of networks are shown in color-code based on Fiedler vectors of $\mathcal{L}$ which is the eigenvector corresponding to the spectral gap; before the transition point networks are decoupled and after transition point they are coupled \cite{Radicchi2013NatPhys}.
}
\label{SD}
\end{figure}

De Domenico et al.~\cite{DeDomenico2016NatPhys} numerically tested for the super-diffusion phenomena on multiplex networks composed of two layers which first and second layers are Erd\H{o}s-R\'enyi random graphs from $\mathcal{G}(N,p_1)$ and $\mathcal{G}(N,p_2)$ respectively, in which, $\mathcal{G}(N,p)$ is the classical random graph ensemble which there are $N$~nodes and each pair of the nodes are connected with probability~$p$. They found out that super-diffusion in these specific multiplexes happens mostly in the same connection probability regime ($p_1 \approx p_2$)~\cite{DeDomenico2016NatPhys}. For a multiplex that its layers are generated with very different probabilities, multiplexity or having multilayer structure has no benefits in terms of diffusion timescale and hence, diffusion in at least one of the layers when considered independently will be faster than this connected multiplex. Fig.~\ref{SD} shows the phenomena of super-diffusion and its phase diagram on random multiplex networks.

While the super-diffusion phase diagram achieved using numerical test, the following question is legitimate: is it possible to design a machine learning algorithm to detect presence of super-diffusion given a multiplex network? To answer this question we propose a supervised ML algorithm.

\paragraph{Deep Learning.} Machine learning (ML), particularly deep learning is now the standard tool in many different areas: from computer vision ~\cite{lecun1998gradient, goodfellow2016deep, krizhevsky2012imagenet, simonyan2014very, szegedy2015going,he2016deep}  to condensed matter physics 
~\cite{Carrasquilla2017NatPhys, vanNieuwenburg2017NatPhys, carleo2017solving, Chng2017PRX, Wetzel2017PRE, Wenjian2017PRE} and quantum systems~\cite{Torlai2018NatPhys, Biamonte2017Nature, Deng2017PRX, Zhang2017PRL}. Application of complex networks to different ML areas has a long story for supervised learning ~\cite{Bertini2011}, unsupervised learning ~\cite{Karypis1999}, semi-supervised learning~\cite{Chapelle2006}. Instead of using complex networks to study deep neural networks, in this paper we present the converse: i.e.~we apply ML methods to study complex networks.  

Which neural network structures are most suited to predict super-diffusion in multiplex networks? To this end the following architectures of Artificial Neural Networks (ANN) were applied: (1) fully connected neural networks (FCNN) with $l_2$ regularization; (2) FCNN with dropout and (3) convolutional neural networks (CNN). The realization was implemented using Keras API~\cite{Keras} backed by TensorFlow~\cite{tensorflow2015-whitepaper}.

\paragraph{Training Set.} We consider duplexes---multiplex networks with two layers (see appendix 2 for the generalization of the framework for multiplex networks with more than two layers). Both the first and second layers  have exactly $N$~nodes and are created independently according to the Erd\H{o}s-R\'enyi random graph model: $\mathcal{G}(N,p_1)$ and $\mathcal{G}(N,p_2)$ respectively. In the $\mathcal{G}(N, p)$ model, a graph is constructed (here each graph forms one duplex layer) by connecting nodes randomly. Each edge is included in the graph with probability $p$ independent from every other edge. Equivalently, all graphs with $N$ nodes and $M$ edges have equal probability of
\begin{equation}
p^M (1-p)^{\binom{N}{2}-M}.
\end{equation}
Here the parameter $p$ in this model can be thought of as a weighting function---as $p$ increases from 0 to 1, the model becomes more and more likely to include graphs with increasing numbers of edges and increasingly less likely to include graphs with fewer edges.  In particular, the case $p = 1/2$ corresponds to the case where all $2^{\binom{N}{2}}$ graphs on $N$ nodes are chosen with equal probability.

Both layers are undirected and unweighted graphs. Connection patterns are encoded in an adjacency matrix $\mathcal{A}^{[1]}~(\mathcal{A}^{[2]})$ for the first (second) layer, where $\mathcal{A}^{[1]}_{ij}=1$ (and then $\mathcal{A}^{[2]}_{ij}=1$) if there is a link between node~$i$ and node~$j$ (all intra-layer links have weights equal to 1). Then we connect these two identical layers by one-to-one interlayer links each with non-negative weight $w$. According to~\cite{GomezSupDiff2013PRL,Radicchi2013NatPhys}, changing the interlayer weights enables one to explore monoplex-multiplex regimes.

We divide the $p_1$-$p_2$ phase diagram into bins with $\delta p=0.02$ for the training phase and $\delta p=0.01$ in the testing phase. For each bin, we create $30$ training and $20$ test sample multiplex networks $(\mathcal{G}(N,p_1),\mathcal{G}(N,p_2))$. If super-diffusion will be exhibited  
\begin{equation}
\lambda_2(\mathcal{L}^{sup}) \geq \text{Max}\{ \lambda_2(\mathcal{L}_1), \lambda_2(\mathcal{L}_2)\}
\end{equation}
then we label the input set $\{ \mathcal{A}^{[k]}, \mathcal{A}^{[m]},\dots \}$ as 1, and otherwise with 0.
\newline
We generated $75,000$ random graph instances for the training set and $200,000$ instances for the test set.

\paragraph{Fully connected neural networks for super-diffusion prediction.} To feed our data to fully connected neural network we make the following transformation. Since adjacency matrices are symmetric $N\times N$ with zeros on the diagonal, we can represent multiplex networks as a matrix with an upper triangle equal to the upper triangle of the first layer and a lower triangle equal to the lower triangle of the second layer. Then this matrix can be reshaped to a $N^2\times 1$ column matrix.

To test the quality recovered from use of a fully connected neural network (FCNN) we choose a simple architecture which consists of one input layer, one hidden layer and one output layer with two neurons. The activation function of the hidden layer is the so-called, ReLU (Rectified Linear Unit) function. To use our FCNN as a classifier, the output layer is activated by a sigmoid function. 
For the first FCNN model, the loss function was chosen as average cross-entropy between predicted and actual values with additional $l_2$ regularization term to prevent over-fitting. The optimization procedure of the loss function was done by the extension to stochastic gradient descent~\cite{DBLP:journals/corr/KingmaB14}. 

The resulting accuracy of the model is  $93\%$ on the test data. The FCNN with dropout recovers the same result.

\begin{figure}[H]
    \includegraphics[width=0.8\textwidth]{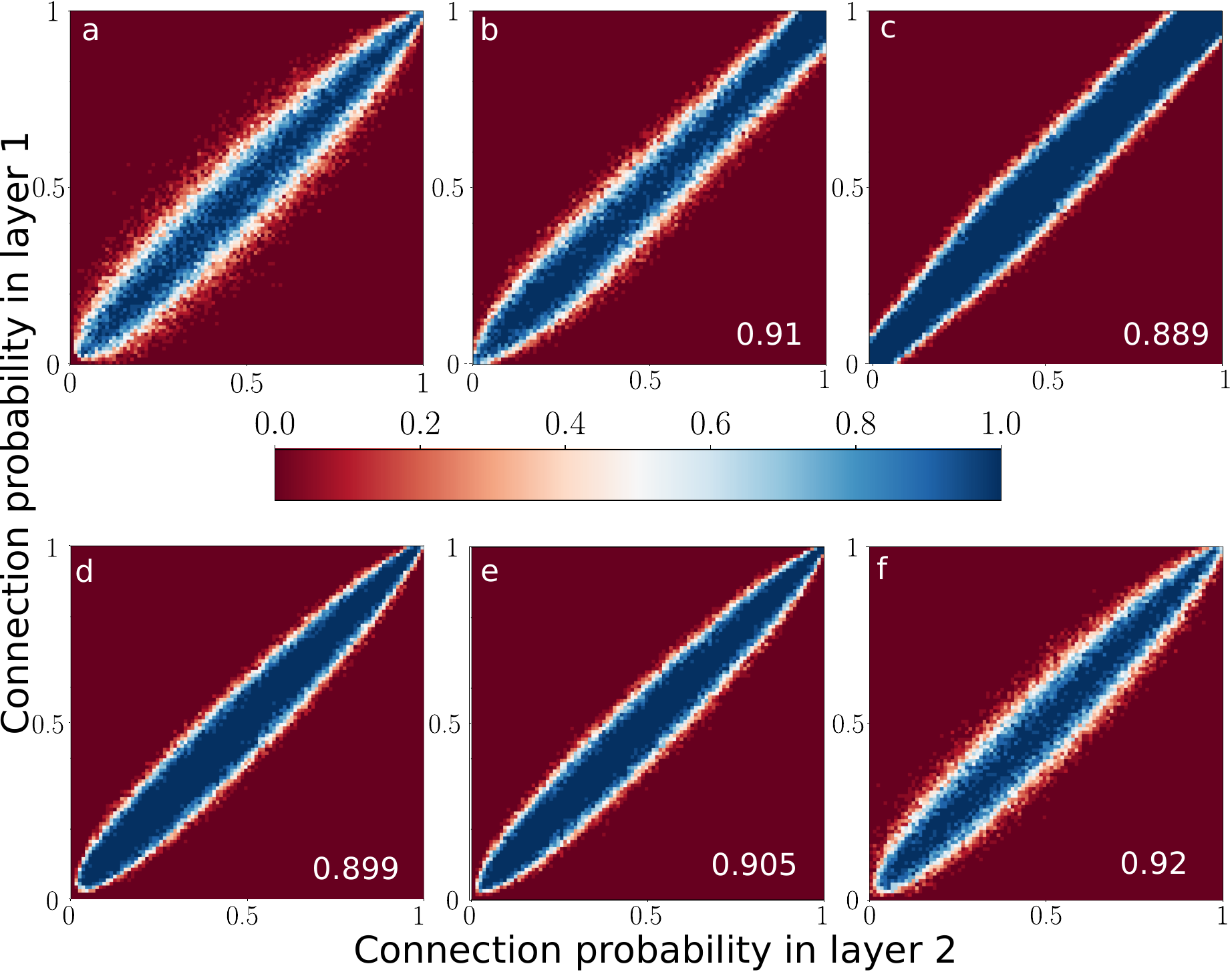}
    \caption{({\bf Machine learning super-diffusion.}) Heatmap formed from averaging of ideal super-diffusion over (a) test data and predictions with (b) FCN with dropout, (c) FCN with l2 regularization, (d) CNN at first, (e) at fifth (f) and at twentieth epoch . Those heatmaps are just for data visualization, each p1-p2 square (with $\delta p = 0.01$) contains the percentage of super-diffusion classified instances, The Structural Similarity with the ground truth (a) are 91\% (b), 88.9\% (c), 89.9\% (d), 90.5\% (e) and 92\% (f).} 
    \label{Plots}

\end{figure}

To reconstruct Fig.~\ref{SD} we plot the distribution of the predicted result. The results are presented in Fig.~\ref{Plots} (b) and (c). We notice that the shape of the distributions varies for real super-diffusion (a) particularly for the left-down and up-right corners of the plot.

\begin{figure*}[ht]
	\includegraphics[width=1\textwidth]{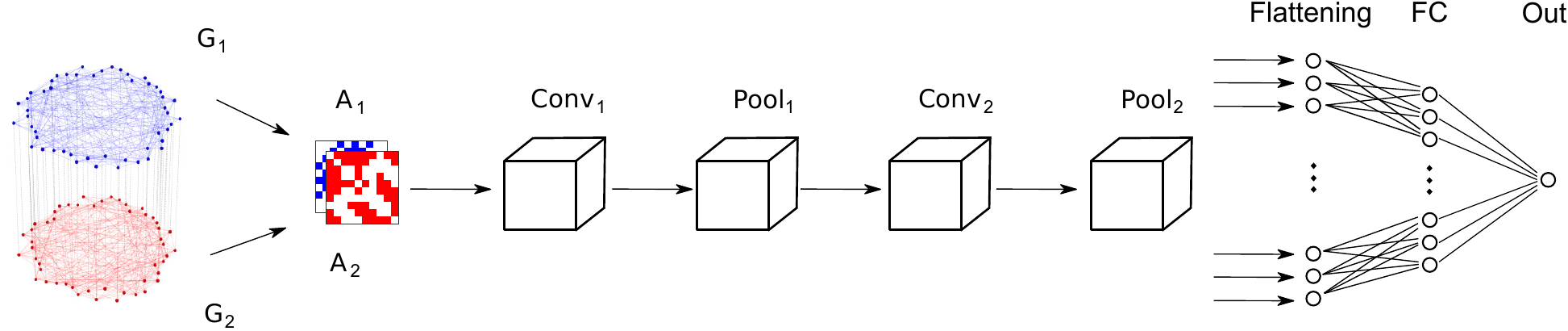}
	\caption{\label{CNN} ({\bf Deep neural network architecture detecting super-diffusion.}) Two-layer random multiplex instances were randomly generated; each layer was generated independently from random graph model $\mathcal{G}(N,p)$, which $N$ is the number of nodes ($=50$) and $p$ is the connection probability. Adjacency matrices of each layer is represented as binary images of $50\times 50$ and then fed to the CNN as two separated channels ($\mathcal{A}^{[1]}$, $\mathcal{A}^{[2]}$) (the complete input shape is [$50\times 50\times 2$]). A first convolution (CONV1) is performed using a $5\times 5$ filter, same padding, ReLU activation function and 32 channels (the output shape is [$50\times 50\times 32$]). One max pooling layer (POOL1) is used to half the input shape with window size of $2\times 2$ and same padding (resulting in the shape of [$25\times 25\times 32$]). The second convolution (CONV2) has the same properties of (CONV1) except for the number of channels---growing now to 64 (shape of [$25\times 25\times 64$]). After a second max pooling layer (POOL2) that has the same properties of (POOL1) the resulting data (of shape [$13\times 13\times 64$]) is flattened to a single vector (of [10816] elements) and fed to a fully connected layer (of [1024] nodes) and to an final output layer (with [1] output node). The  output is as a binary classifier which determines if the instance exhibits super diffusion or not. 	}
\end{figure*}

\paragraph{Convolutional neural networks for super-diffusion prediction.} The design of convolutional neural networks (CNN) which we utilized for the problem is shown on Fig. \ref{CNN}. The input of the CNN is two channel $50\times50$ binary image followed by two convolutional layers with a small size $5\times5$ filters and pooling layers. The result is flattened and connected to FCN with two outputs with preceding activation sigmoid function. 

To feed data to the CNN we represented the 2 separated adjacency matrices of a multiplex network as binary images of 50 by 50 pixels and assigned those as two separated channels on the input layer. This procedure is shown on the left side of Fig. \ref{CNN}. 

Interestingly enough the resulting accuracy $94\%$ is slightly better for the CNN then for FCNN and the shape of the distribution of predicted result looks more similar to the real distribution Fig.~\ref{Plots}(d-f). This fact points out that the representation of a multiplex network as a multi-channel image for CNNs inputs is quite promising feature to investigate properties of multiplexes. And probably more complicated CNN's architectures could perform better results. See appendix 1 for more information on the accuracy of the results predicted by CNN.

\paragraph{Conclusions.} We have found that convolutional  neural network technology, developed for applications such as computer vision, can be used to detect super-diffusion in multiplex networks. We argue that this idea is quite natural and likely can attribute some of its success due to our representation of a multiplex layers adjacency matrix as channels of an image. 

While traditional machine learning techniques have long been part of the complex networks tool-kit, deep learning is in its nascent days of application.  Indeed, two-layer FCN was used as perhaps the simplest example of deep neural network, likewise with our simplistic simple CNN structure ($l_2$ regularization and drop-out techniques were used to prevent over fitting).  In both cases, training resulted in high detection probability, thereby confirming proof of principle even for simplistic neural network architectures (over $94\%$ accuracy). 

Based on our findings, we however anticipate a further use of deep learning in the field of complex networks, such as detecting phase transitions, and network dismantling. Interestingly, the application of complex networks as a tool to study properties of supervised learning ~\cite{Bertini2011}, unsupervised learning ~\cite{Karypis1999} and semi-supervised learning~\cite{Chapelle2006} has long been considered. By applying deep learning to study properties of complex networks we then hope to form a bridge and a future where these two subjects compliment each other with deep learning regularly applied to complex networks. 

As mentioned, the main property of multiplex networks which appears useful and quite natural for CNN, is the adjacency matrix which can be represented as data-set for every layer. This has the same format as image channels and is likely accountable as to why CNN filters can detect features responsible for super-diffusion in multiplex networks so readily.\\

\begin{figure*}
\RawFloats
\centering

\sbox0{%
  \begin{minipage}[b]{.59\textwidth}
  \includegraphics[width=10.5cm,]{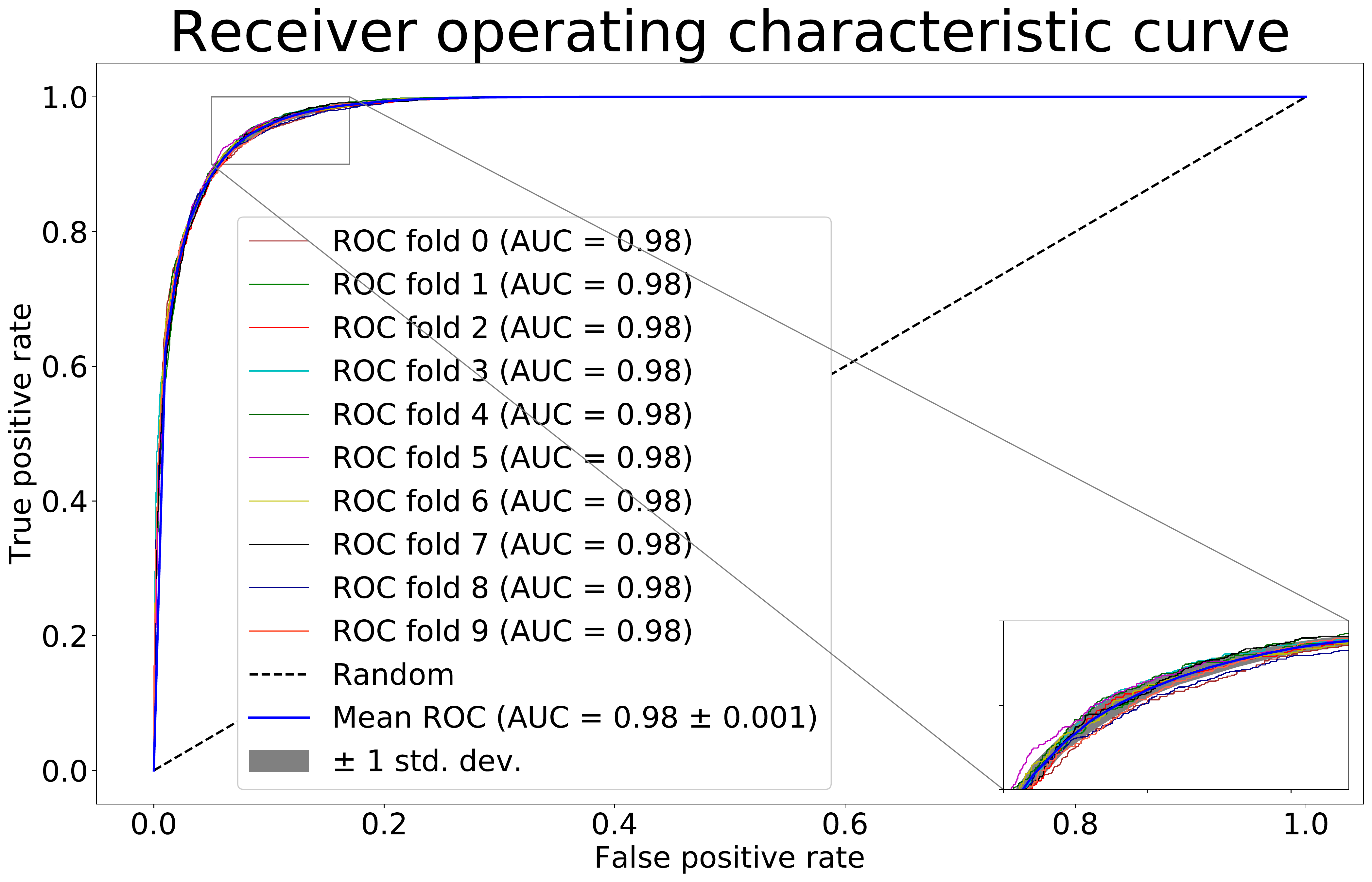}
  
  \vfill
  \caption{\label{roc}({\bf Performance of SD detection (ROC curves)}) Cross-validation on a stratified 10-fold data. Two datasets with different partitions of the connection probabilities of 75K and 36K elements are merged and shuffled. Then, the entire dataset is split into 10 folds, with each fold containing the same percentage of SD cases (17\%).
The architecture is initialized 10 times, with the training occurring on 9 folds and testing being performed on the remaining 10th fold. 
Areas under the curve (AUC) for different folds and the mean AUC are shown. 
}
  \end{minipage}}
\usebox{0}\hfill
\begin{minipage}[b][\ht0][s]{.4\textwidth}
\begin{minipage}{\textwidth}
\centering

\includegraphics[width=8cm]{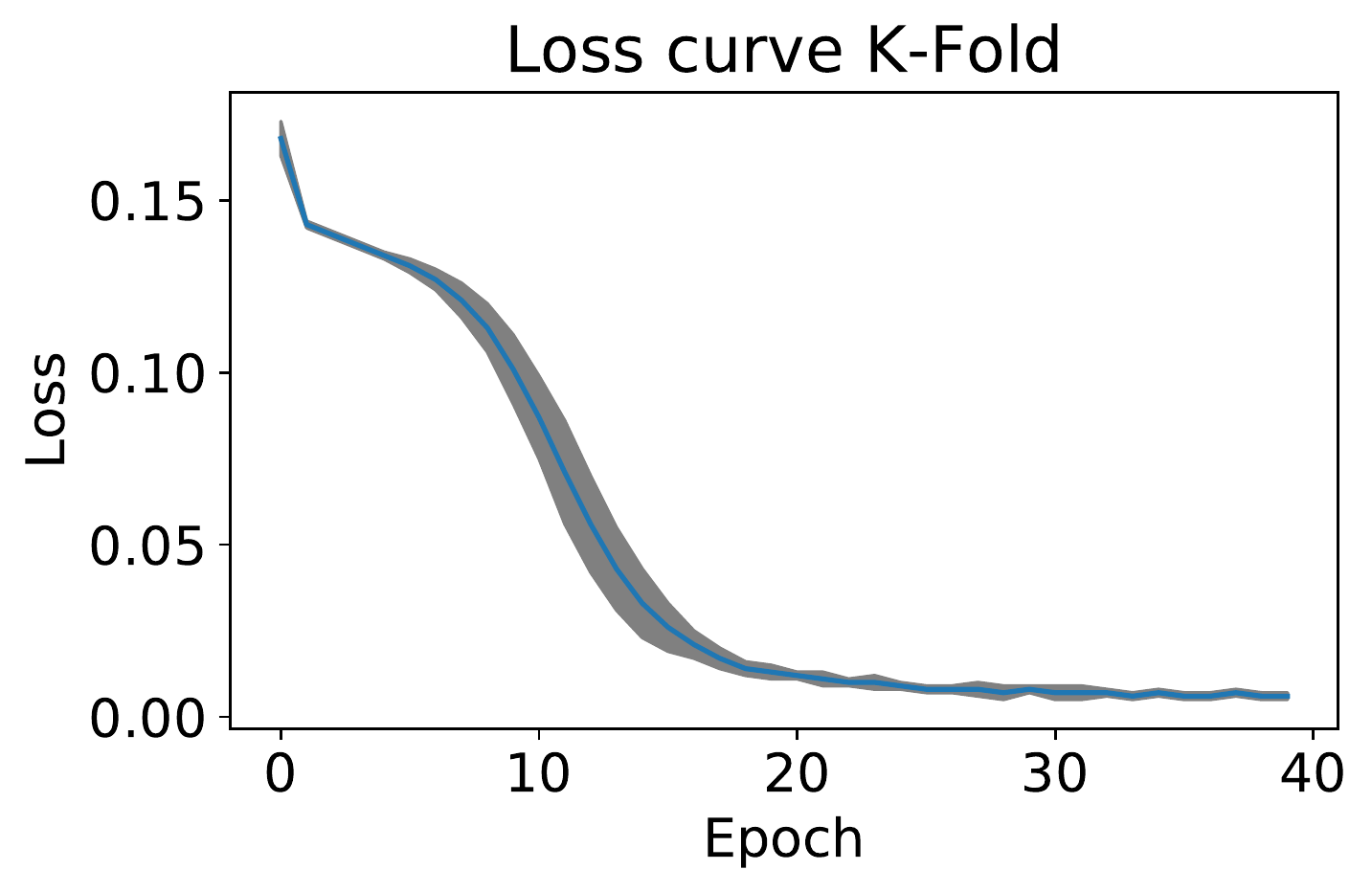}

\caption{\label{loss}({\bf Loss curve for K-Fold validation}) The binary cross-entropy was used as a loss function. Shown are the mean value (solid line) and the standard deviation (shaded).
}

\captionof{table}{MSE and SSIM metrics as a function of learning epochs.}
\medskip
\label{mse_sim}
\begin{tabular}{c|c|c}
        Epoch  & MSE & SSIM \\
        \cline{1-3}
            1 & 0.028 & 0.894\\
            5 & 0.023 & 0.901\\
            10 & 0.018 & 0.911\\
            15 & 0.012 & 0.922\\
            20 & 0.012 & 0.922\\
            25 & 0.029 & 0.894\\
            30 & 0.019 & 0.904\\
            35 & 0.012 & 0.922\\
            40 & 0.014 & 0.919\\
     \end{tabular}\\
\end{minipage}
\vfill
\end{minipage}

\end{figure*}

\paragraph{Appendix 1}
Here we show some metrics obtained by a cross validation where we divided the whole set in 10 stratified folds (each preserve the percentage of samples for both classes). For every step we kept 9 folds for training and 1 for testing.  Fig.~\ref{roc} shows the ROC curve (receiver operating characteristic curve) of the different folds and the mean AUC  (area under the ROC curve) of 98\%.
The average accuracy obtained from the different folds is 94\%. Table~\ref{conf_matrix} shows the confusion matrices for K-Fold validation (top) and for the independent test set (bottom).
Fig~\ref{loss} shows the loss curve for the 10-fold stratification over 40 epochs, as additional metrics we generated heatmaps for each epoch and measured against the ground truth with Mean Squared Error and Structural Similarity Index with results shown in Table \ref{mse_sim}.

\begin{figure*}[ht]
    \includegraphics[width=1\textwidth]{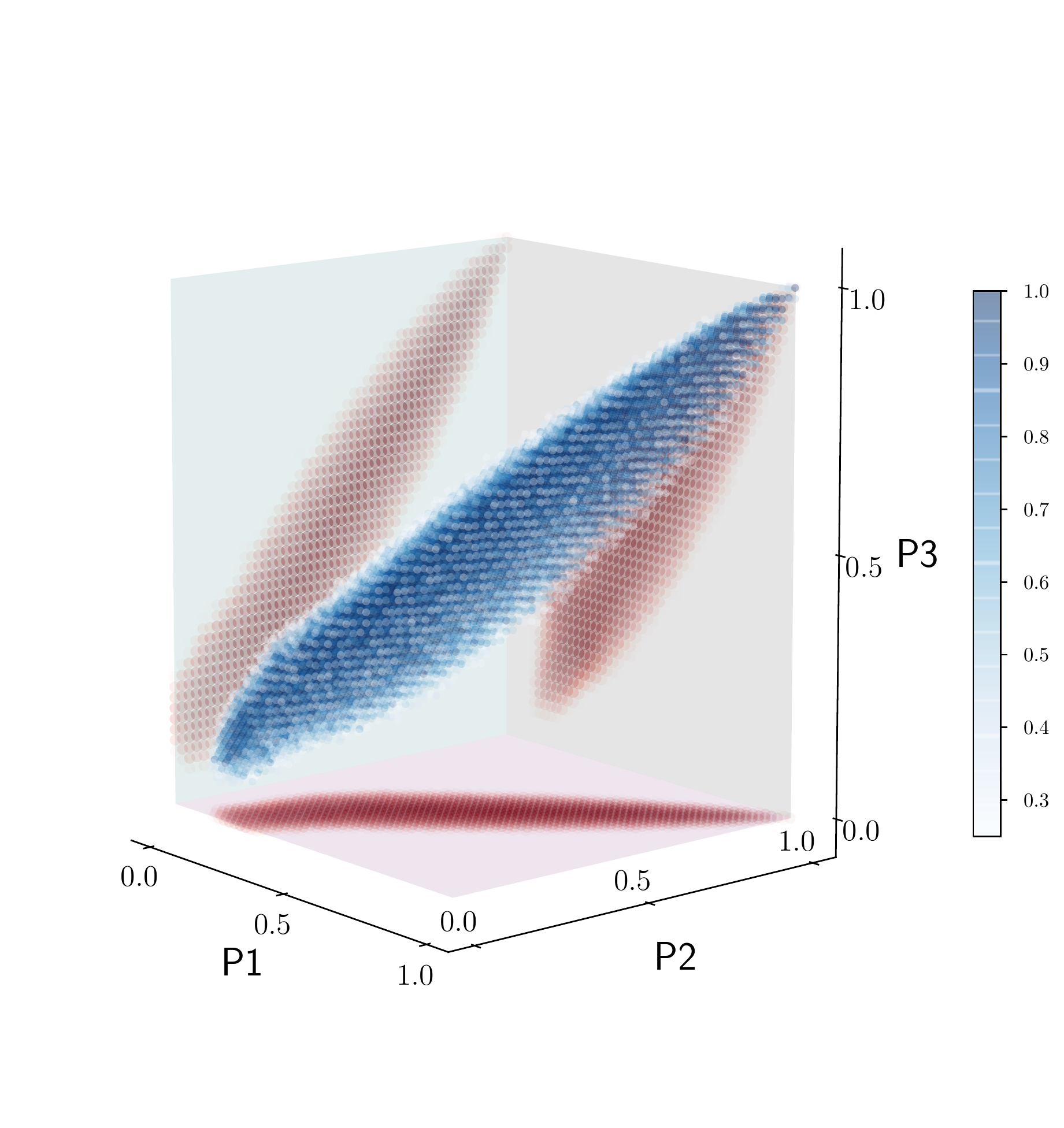}
    \caption{({\bf Generalizing to more than two layers.}) Predicted super diffusion phase diagram in 3-layer random multiplex networks. Each layer is an Erd\H{o}s-R\'enyi random graph sampled from $\mathcal{G}(N,p)$ ensemble, where $N$ is number of  nodes ($N=50$) and $p$ is the connection probability. For each ($p_1,p_2, p_3$) we create 20 samples of the 3-layer random multiplex networks, then we check the condition of super-diffusion on each of them. Color coding is based on the probability of the existence of super diffusion in each bin. For clarity we kept only data points corresponding to probabilities higher than 0.5. Projections of the main heatmap in (x,y), (x,z) and (y,z) planes represent the super diffusion in corresponding sub-duplexes formed by different combinations of layers. The only change in the DNN architecture is the input layer which is composed of three channels (the complete input shape is [50 x 50 x 3]). The classification accuracy obtained is 94\% } 
    \label{3layer}
\end{figure*}

\paragraph{Appendix 2} Here we extend the framework presented in the main text to 3-layer multiplex networks (feeding CNN with adjacency matrices of the layers and following training procedure as explained in the main text). Fig.~\ref{3layer} shows the ability of our deep learning framework to predict super diffusion in case of multiplex networks with more than two layers. This proves the extension feasibility of the machinery to higher order multiplex networks.
In the case of multiplex networks with three layers, We divide the $p_1$-$p_2$-$p_3$ phase diagram into bins with $\delta p=0.020$. For each bin we create $15$ training samples multiplex networks $(\mathcal{G}(N,p_1),\mathcal{G}(N,p_2),\mathcal{G}(N,p_3))$. If super-diffusion will be exhibited  
\begin{equation}
\lambda_2(\mathcal{L}^{sup}) \geq \text{Max}\{ \lambda_2(\mathcal{L}_1), \lambda_2(\mathcal{L}_2), \lambda_2(\mathcal{L}_3)\}
\end{equation}
then we label the input set $\{ \mathcal{A}^{[k]}, \mathcal{A}^{[m]},\dots \}$ as 1, and otherwise with 0.
\\
We generated $100,000$ random graph instances for the training set and $50,000$ instances for the test set (the latter has same $\delta p=0.020$ and $20$ samples per bin). Fig \ref{3layer} shows the resulting heatmap on test set.

\paragraph{Appendix 3} Here we extend the framework presented in the main text to 2-layer multiplex networks of arbitrary $N$ nodes. We found that the original architecture can be extended to correctly classify bigger multiplex by stacking successive convolution (Conv) and pooling (Pool) layers. In the original $N=50$ size we stacked two Conv plus Pool layers, the only change in the DNN architecture is the number of Conv plus Pool layers, The number of such stacks is $\lceil N \bmod 50 \rceil +1$.
Each layer is an Erd\H{o}s-R\'enyi random graph sampled from $\mathcal{G}(N,p)$ ensemble, where $N$ is number of  nodes ($N=50, 100, 150, 200, 250, 300, 350, 400, 450 $ and $ 500$) and $p$ is the connection probability. Likewise the $N = 50$ case, we  divide the training set in bins of $\delta p= 0.002$ and the test set with $\delta p=0.001$, we create 30 samples per bin in train and 20 in test sets.

\paragraph*{\bf Data availability.}
The data that support the plots within this paper and other findings of this study as well as all source code are available online.\footnote{https://github.com/vitomichele/Repository-of-paper-Deep-Learning-Super-Diffusion-in-Multiplex-Networks}

\twocolumngrid

\bibliographystyle{unsrt}
\bibliography{refs}  

\onecolumngrid
\appendix

\newpage

\onecolumngrid 

\newpage
\captionof{table}{Confusion Matrix for 10-fold cross-validation (top) and for independent test set (bottom).}
\label{conf_matrix}
\begin{tabular}{l|l|c|c|c}
\multicolumn{2}{c}{}&\multicolumn{2}{c}{}&\\
\cline{3-4}
\multicolumn{2}{c|}{}&Positive&Negative&\multicolumn{1}{c}{}\\
\cline{2-4}
& Positive & $1602.1\quad$ & $291.7\quad$ & PPV = $0.85\pm0.04$\\
\cline{2-4}
& Negative & $385.9\quad$ & $8819.3\quad$ & NPV = $0.96\pm0.01$\\
\cline{2-4}
\multicolumn{5}{c}{} \\
\multicolumn{1}{c}{} & \multicolumn{1}{c}{} & \multicolumn{1}{c}{TPR = $0.81\pm0.06$} & \multicolumn{1}{c}{TNR = $0.97\pm0.01$}\\
\multicolumn{1}{c}{} & \multicolumn{1}{c}{} & \multicolumn{1}{c}{(Sensitivity)} & \multicolumn{1}{c}{(Specificity)}\\
\end{tabular}
\newline
\newline

\begin{tabular}{l|l|c|c|c}
\multicolumn{2}{c}{}&\multicolumn{2}{c}{}&\\
\cline{3-4}
\multicolumn{2}{c|}{}&Positive&Negative&\multicolumn{1}{c}{}\\
\cline{2-4}
& Positive & $30193\quad$ & $5494\quad$ & $PPV=0.85$\\
\cline{2-4}
& Negative & $6772\quad$ & $157541\quad$ & $NPV=0.96$\\
\cline{2-4}
\multicolumn{5}{c}{} \\
\multicolumn{1}{c}{} & \multicolumn{1}{c}{} & \multicolumn{1}{c}{\hspace*{4mm} TPR = $0.82 \hspace*{5mm}$} & \multicolumn{1}{c}{
\hspace*{4mm} TNR = $0.97$ \hspace*{5mm}}\\
\multicolumn{1}{c}{} & \multicolumn{1}{c}{} & \multicolumn{1}{c}{(Sensitivity)} & \multicolumn{1}{c}{(Specificity)}\\
\end{tabular}

\clearpage

\newpage

\end{document}